# Objective Measurement of AI Literacy:
# Development and Validation of the AI Competency Objective Scale (AICOS)

Authors:

André Markus[a]*, Astrid Carolus[b] , and Carolin Wienrich[a]

[a] Psychology of Intelligent Interactive Systems, Institute Human-Computer-Media, Julius-Maximilians-University, EmilFischer-Straße 50, 97074 Wuerzburg, Germany

[b] Media Psychology, Institute Human-Computer-Media, Julius-Maximilians-University, Oswald-Külpe-Weg 82, 97074 Wuerzburg, Germany

*Correspondence: andre.markus@uni-wuerzburg.de

## Abstract

As Artificial Intelligence (AI) becomes increasingly pervasive in everyday life, AI literacy is widely recognized as a set of essential competencies for navigating AI-driven environments safely, responsibly, and effectively. There is a growing need to assess this construct, for example, to inform targeted educational interventions. Although several measurement tools already exist, many show limitations regarding subjective data collection methods, differentiation between target groups, validity, and the integration of recent developments, such as Generative AI literacy. To address these limitations, this study introduces the AI Competency Objective Scale (AICOS) - an instrument grounded in a competency-oriented framework that enables a more objective assessment of AI literacy sub-competencies. AICOS draws on established theoretical models, integrates validated items from prior instruments, and explicitly incorporates Generative AI literacy as a distinct dimension. The AICOS provides a sound and comprehensive measure of AI literacy, and initial analyses indicate the potential for a modular structure. A preliminary short version of the scale has also been developed. Due to its methodological foundation, extensive validation, and incorporation of recent technological advancements, the test represents a valuable tool for scientific research and practical applications in educational and professional contexts. The AICOS contributes to the standardization of AI literacy assessment and supports the targeted development of AI-related competencies across diverse populations.

# 1 INTRODUCTION

The proliferation of artificial intelligence (AI) is transforming both the professional and personal spheres. In the workplace, AI is optimizing processes in healthcare, such as diagnostics, medical imaging, and surgery (Benzakour et al., 2023; Kumar et al., 2023; Larrazabal et al., 2020), and in education, where it is improving performance assessments and enabling personalized learning (Chiu & Silva, 2024; Pratama et al., 2023). AI is also used in finance and science for applications such as fraud prevention (Khare & Srivastava, 2023) and innovative hypothesis generation (Sourati & Evans, 2023; Xu et al., 2021). Changes in the workplace are increasing the demand for AI-related skills (Peede & Stops, 2024; Verma et al., 2022), and the importance of AI skills is also growing in the private sphere. In daily life, AI powers smart appliances, personalized recommendation systems, and app-based technologies such as face-swapping and AI-generated voice clones. At the same time, the growth of deepfakes raises concerns about disinformation, especially in political campaigns (Khanjani et al., 2023; Kharvi, 2024). With the introduction of easy-to-use AI tools like ChatGPT, AI has become more accessible to a wide audience. The European Parliament (2024) has responded to the proliferation of AI in the professional and private spheres with the AI Act. This law regulates the use of AI systems in Europe and includes requirements for developing AI skills. For example, Article 4 of the AI Act requires providers and operators of AI systems to ensure that their employees and contractors have sufficient AI skills. Against this backdrop, valid measurement of AI skills is becoming increasingly important in systematically assessing them, identifying deficits, and developing targeted training programs. So far, subjective self-assessments have predominated, although these depend heavily on self-assessment skills and are subject to the person's own biases (Chiu et al., 2024). By contrast, objective tests are rarely available. The present study addresses this need by developing a scientifically valid scale with an objective approach to measuring AI literacy. It builds on existing measurement tools, overcomes their limitations, integrates current competency models and developments in AI, and enables differentiated measurement across heterogeneous populations. A first version of a shorter version of this scale is also being developed, promoting a more economical and efficient use of the instrument. Thus, the AI Competency Objective Scale (AICOS) is taking a decisive step toward assessing AI skills at different levels and identifying training needs. This study employs the term AI literacy to emphasize its broad educational relevance and multidimensional nature, as the construct is commonly understood to comprise multiple sub-competencies (Long & Magerko, 2020). AICOS builds on this understanding by applying a competency-oriented framework to operationalize and assess these competencies in alignment with established competency structures and psychometric models (Carolus, Koch, et al., 2023).

## 1.1 Definition of AI Literacy

AI literacy is defined as the set of skills that enable individuals to evaluate, interact with, and utilize AI technologies effectively and meaningfully in various aspects of life (Long & Magerko, 2020). Based on Ng et al. (2021), Carolus, Koch, et al. (2023) developed the Meta AI Literacy Scale (MAILS), a robust and widely cited subjective assessment tool. It conceptualizes AI literacy along two dimensions:

First, **core AI literacy** competencies include *"Know and Understand AI," "Use and Apply AI," "Detect AI," "Evaluate and Create AI," and "AI Ethics"* (see Table 1). This structure aligns with Bloom's taxonomy of educational objectives (Bloom, 1956), which categorizes cognitive skills from basic recall to higher-order thinking, such as analysis and creation. However, unlike Bloom's strict hierarchy, MAILS — and our approach — conceptualizes these competencies as interconnected rather than strictly sequential. For example, individuals may be able to design AI systems without yet being able to critically assess their ethical implications. Second, **AI self-management** encompasses psychological factors such as *AI Self-efficacy* and *AI Self-Competency* in dealing with AI. The modular design of the scale enables flexible use across various contexts, with each subscale validated and independently applicable. Moreover, Generative AI literacy has recently emerged as a distinct facet that challenges and extends traditional definitions of AI literacy (Annapureddy et al., 2024; Zhao et al., 2024). While Bloom's taxonomy remains valuable, generative AI literacy demands an integrative approach combining theory, practice, and critical reflection. Its content-generation capability from minimal input is rarely covered in existing competence models, highlighting the need for adaptation (Annapureddy et al., 2024; Jin et al., 2024) (Table 1). Overall, the definition of AI literacy underscores its growing importance in education and the workplace, emphasizing the need for valid and objective assessment tools. Despite the increasing integration of AI literacy content in curricula and training, many individuals still lack essential competencies (Antonenko & Abramowitz, 2023; Carolus, Augustin, et al., 2023; Cetindamar et al., 2022; Markus et al., 2024; Mertala & Fagerlund, 2024; Nader et al., 2024; Yim, 2024). Reliable measurement instruments are essential for identifying gaps and providing support for evidence-based instruction.

**Table 1**

*Sub-competencies of AI literacy, as defined by Carolus, Koch et al. (2023), and Annapureddy et al. (2024), mapped to Bloom's Taxonomy.*

| Cognitive Level of Bloom's Taxonomy | Short Definitions of Subcompetencies of AI Literacy (Term within this study) | Bloom-level assignment, according to Carolus, Koch et al. (2023) |
|---|---|---|
| 6) Create | **Evaluate and Create AI (Create AI)**<br>Higher cognitive skills include evaluating, assessing, predicting, and developing AI applications. | 5 + 6 |
| 5) Evaluate | **Detect AI (Detect AI)**<br>Recognize if an application is based on AI or not | 5 |
| 4) Analyse | **AI Ethics (AI Ethics)**<br>Consideration of human values such as fairness, responsibility, transparency, ethics, and safety in the context of AI. | 4 |
| 3) Apply | **Use and Apply AI (Apply AI)**<br>Apply knowledge, concepts, and AI technologies in different scenarios. | 3 |
| 2) Understand | **Know and Understand AI (Understand AI)**<br>Understand the basic functions of AI and how to use AI applications. | 1 + 2 |
| 1) Remember | **Generative AI Literacy (Generative AI)**<br>Understand, reflect upon, and competently apply generative AI. | * |

*Note.* While Bloom's taxonomy traditionally follows a strict hierarchical order, the competencies in AICOS are conceptualized as interconnected and non-linear in their development. *Generative AI Literacy is a novel and integrative competency that requires further research to clarify its placement within or beyond Bloom's framework.

## 1.2 Measures of AI literacy

AI literacy scales provide a framework for assessing the sub-competencies of AI literacy, identifying deficits, designing targeted interventions, and evaluating their effectiveness. Measurement tools help identify AI skill gaps across different populations, ensuring equal opportunities for developing the knowledge and skills needed in an AI-driven society. Many existing AI literacy scales rely on self-reporting, capturing only individuals' subjective perceptions of their AI knowledge and skills (Carolus, Koch, et al., 2023; Laupichler et al., 2023; Lee & Park, 2024; Ng et al., 2024; Pinski & Benlian, 2023; Wang & Chuang, 2024; Zhang et al., 2024). Chiu et al. (2024) argue that subjective measures reflect perceived rather than actual AI literacy, and self-perception often deviates from real ability (Tempelaar et al., 2020; Weber et al., 2023). Other empirical studies support this hypothesis. Moorman et al. (2004) demonstrate that objective knowledge reflects a person's actual knowledge, whereas subjective knowledge only reflects what people think they know. In this context, Raju et al. (1995) describe subjective knowledge as the "feeling of knowing". Research indicates that subjective and objective knowledge is often uncorrelated or only weakly correlated, a pattern also observed in the relationship between subjective and objective AI literacy (Carlson et al., 2009; Dunning, 2011; Klerck & Sweeney, 2007; Ma & Chen, 2023; Moore & Healy, 2008; Pinski et al., 2024; Waters et al., 2018). To objectively measure AI literacy, alternative methods are needed to assess actual AI literacy, which is critical for evaluating training efforts, determining training needs, or even making evidence-based personnel decisions.

#### *1.2.1 Objective AI Literacy Measures*

Recent research has investigated the objective measurement of AI literacy using knowledge-based multiple-choice questions (MCQs). As a fundamental competency indicator, knowledge lends itself to assessment through MCQs, which ensure consistent questions, unambiguous correct responses, and reduced scorer subjectivity (Brady, 2005; Hammond et al., 1998; Sternberg & Horvath, 1995). Consequently, MCQs are often regarded as a more objective measurement approach, although this objectivity is not absolute, as validity depends on factors such as item quality, test design, and test-taking strategies (Embretson & Reise, 2000; Messick, 1995). The present study identified several knowledge-based measures as an objective approach to measuring AI literacy using MCQs (see Table 2 or Supplementary Material 1). While many of these methods have high content validity (often confirmed by external expert ratings), they have weaknesses in important test-theoretical quality criteria. In particular, sample representativeness is limited by size and the focus on specific target groups (e.g., children or teachers) (Ding et al., 2024; Weber et al., 2023; Yau et al., 2022). In addition, comprehensive analyses of convergent, discriminant, and criterion validity, crucial for psychometric quality, are often missing. Among the instruments identified, those by Hornberger et al. (2023), Chiu et al. (2024), and Pinski et al. (2023) appear particularly promising in terms of their quality and statistical validity:

Hornberger et al. (2023) developed a knowledge-based scale with 30 MCQs and a sorting task based on the conceptualization of Long and Magerko (2020). The scale has been validated in German and shows high structural validity, internal consistency, and convergent validity. However, the validation is based solely on a student sample, which limits its representativeness and lacks evidence of discriminant and criterion validity according to current scientific standards (Cohen, 1992; Peers, 2006). Pinski et al. (2023) also developed a knowledge scale with binary MCQs (two response options). Although the scale is content valid, with evidence of convergent and discriminant validity, it is based on a small sample of non-experts in AI, which limits the generalizability of the results. Additionally, the scale is only available in English, and there are no indications that criterion validity analyses have been conducted. Chiu et al. (2024) developed a scale based on the curriculum of Chiu et al. (2021) and aligned it with guidelines for AI education. It consists of 25 English MCQs with four response options and has satisfactory reliability and construct validity. Despite the large sample size ($N$ = 2380), only 7th and 8th-grade students in Hong Kong were assessed, and convergent and discriminant validity analyses were missing. Despite the remarkable progress made in developing measures of AI literacy, several important scientific challenges remain. These concern both the representativeness of the samples and the fulfillment of all relevant quality criteria of validity. Given these gaps, the present study sets a new benchmark by taking an integrative approach that combines the strengths of existing instruments into a more comprehensive and validated measure.

### 1.3 Present Study

In light of the aforementioned limitations, the present study introduces the AI Competency Objective Scale (AICOS), an instrument based on the model by Carolus, Koch, et al. (2023), which encompasses five sub-competencies and explicitly incorporates Generative AI Literacy (Annapureddy et al., 2024). Rather than developing an isolated new tool, AICOS incorporates content-validated items from established instruments to ensure comprehensive coverage and psychometric rigor. To address the shortcomings of subjective self-assessments, AICOS employs MCQs, enabling more objective measurement and reducing self-report bias. The scale was tested on a heterogeneous sample of varying ages, educational backgrounds, and occupational contexts, ensuring broad applicability across different life stages and professional settings.

This diversity enables the examination of demographic influences on AI literacy, thereby enhancing the instrument's generalizability and practical relevance. Developed within a rigorous quantitative framework, AICOS is validated according to established psychometric standards, including reliability analyses, criterion validity, and convergent and discriminant validity. A key innovation is the explicit inclusion of Generative AI Literacy — a rapidly evolving dimension that is largely absent from existing instruments. Preliminary analyses also assess the feasibility of a shortened version to enable more efficient use without compromising validity. This integrative, methodologically grounded approach positions AICOS as a robust and scalable instrument that addresses the core limitations of earlier tools — namely, limited objectivity, insufficient validation, restricted applicability to target groups, and the neglect of recent technological developments. Ultimately, AICOS offers a scientifically sound and practical instrument for diagnosing AI competencies and informing the design and evaluation of targeted educational interventions.

## 2 METHOD

### 2.1 Item Development

The compilation of items followed a multi-stage scale development process, as outlined by Moosbrugger and Kelava (2020). This process comprised three phases: 1) screening, 2) theory-based categorization and content-driven reduction, and 3) data-supported item analysis.

**Phase 1 (initial item pool).** AICOS employs knowledge-based MCQs to enable efficient large-scale assessment and ensure a standardized measurement of AI literacy, minimizing rater bias (Hammond et al., 1998). To ensure a theoretically grounded and content-valid item pool, the academic literature was systematically screened for existing MCQ-based instruments assessing AI literacy (e.g., Chiu et al., 2024; Hornberger et al., 2023; Kong et al., 2024) (see Table 2 and Supplementary Material 1). In addition, quiz items were sourced from established online learning platforms that promote AI skills, including *Elements of AI* (https://www.elementsofai.com/), *KI-Campus* (https://ki-campus.org/), and the *MOTIV training platform* (https://motiv.professor-x.de/portal/index.html). These platforms were selected because they represent widely used, practice-oriented educational resources that foster foundational AI knowledge among diverse learner groups. The authors reviewed items identified from the screened scales through consensus discussions to ensure their relevance for AICOS's intended adult target population. Items with a strong population-specific bias (e.g., a formulation designed specifically for children) or conceptual misalignment were excluded. Minor wording adjustments were made only when necessary to improve clarity or broaden applicability. This pragmatic approach aimed to harmonize the item pool and ensure its suitability for a diverse adult audience, regardless of educational or professional background. This process resulted in an initial item pool of 282 items. Where necessary, the items were translated into German and then checked for comprehensibility and readability in terms of face validity, in accordance with the guidelines of Moosbrugger and Brandt (2020). Two student assistants then double-checked them.

**Phase 2 (test item pool).** After assembling an initial item pool that exceeded feasible survey length, item reduction was carried out based on multiple criteria. These included qualitative formulation guidelines (e.g., overly complex wording, ambiguity, or technical flaws), psychometric considerations (e.g., violations of assumptions from classical test theory or item response theory), and content redundancies (e.g., substantial overlap between items). The evaluation was guided by established recommendations for item construction and test development (Baker, 2001; Bühner, 2011; Krebs, 2004; Moosbrugger & Kelava, 2020). Detailed criteria for item inclusion and exclusion are outlined in the coding rules (Supplementary Material 2). This evaluation was carried out by two researchers from the field of artificial intelligence (social scientists). Items flagged according to the coding rules were either revised or removed. The review was conducted in two rounds: in the first, each rater independently evaluated the items, resulting in good inter-rater agreement (*Cohen's Kappa* = .77). In the second round, any disagreements were resolved through discussion until full consensus on item inclusion or exclusion was reached. To ensure high content validity, the remaining items were categorized by the raters according to the six sub-competencies of AI literacy defined by Carolus, Koch, et al. (2023) and Annapureddy et al. (2024) (Table 1), achieving good inter-rater agreement (*Cohen's Kappa* = .72). All categorizations were finalized by consensus. To address underrepresented sub-competencies, theory-based items were iteratively developed by the authors based on the models of Long and Magerko (2020), Carolus, Koch, et al. (2023), and Annapureddy et al. (2024) (cf. Supplementary Material 3). This resulted in four items for Create AI, one for Detect AI, five for AI Ethics, eighteen for Generative AI, and two for Understanding AI. Following Moosbrugger and Brandt (2020) and Krebs (2004), two student assistants independently reviewed these items for face validity, assessing their comprehensibility and readability. Subsequently, the same two independent raters who categorized the other items assigned these new items to the six AI literacy sub-competencies used previously, achieving high agreement (*Cohen's Kappa* = .81). The final test item pool consists of 107 items. The psychometric properties of all items, including the newly developed ones, were examined in the comprehensive item response analysis (IRT) described below (Phase 3). The complete item list, along with category labels, is provided in Supplementary Material 3. Category distributions and abbreviations appear in Table 2.

**Phase 3 (final item pool).** The items in the test item pool from Phase 2 were analyzed for their psychometric properties using a suitable IRT model (see Section 2.5). The items were evaluated in a complementary statistical analysis and a content quality check. Based on this analysis, items were retained, revised, or eliminated, resulting in the final item pool. IRT is particularly well-suited for MSQ tests, such as AICOS, as it models the relationship between individual item responses and the latent trait, providing detailed estimates of item difficulty, discrimination, and guessing parameters (Embretson & Reise, 2000; Hambleton et al., 1991). This item-level precision surpasses classical test theory by accounting for item heterogeneity and producing ability estimates that are less sample-dependent (De Ayala, 2013). Additionally, while traditional validation methods assess overall test validity, IRT enhances item selection, scoring accuracy, and interpretability, thereby supporting the development of psychometrically robust and diagnostically informative tests that reliably measure varying levels of AI literacy (Reise & Waller, 2009).

**Table 2**

*Sources and number of remaining test items per instrument, with allocation to respective AI literacy sub-competencies*

| Source | Items (*n*) | Categories (Abbreviation) | Labels in Item Pool | Items (*n*) |
|---|---|---|---|---|
| Chiu et al. (2024) | 12 | Apply AI (AA) | AA01-AA23 | 23 |
| Elements of AI | 5 | Create AI (CA) | CA01-CA12 | 12 |
| Hornberger et al. (2023) | 10 | Detect AI (DA) | DA01-DA10 | 10 |
| Adapted from Annapureddy et al. (2024) | 20 | Ethics AI (EA) | EA01-EA15 | 15 |
| KI Campus | 9 | Generative AI (GA) | GA01-GA19 | 19 |
| Kong et al. (2024) | 7 | Understanding AI (UA) | UA01-UA28 | 28 |
| Melsión et al. (2021) | 1 | | | |
| MOTIV | 12 | | | |
| Ng et al. (2024) | 2 | | | |
| Pinski et al. (2024) | 3 | | | |
| Rodríguez-García et al. (2021) | 6 | | | |
| Self-created | 10 | | | |
| Weber et al. (2023) | 3 | | | |
| Yau et al. (2022) | 3 | | | |
| Zhang et al. (2024) | 4 | | | |

## 2.2 Participants

To ensure a stable estimation of the IRT parameters and account for the test length, a sample size of at least N = 500 was targeted (DeMars, 2010; Embretson & Reise, 2013). Participants were recruited through the Prolific online panel.

Inclusion criteria were residence in Germany and German as the native language. Participation was compensated with 7.20£ per hour. After excluding inappropriate data (*n* = 32; determined via attention checks; see Section 2.3), data from *N* = 514 participants ($n_{males}$ = 273, $n_{females}$ = 235, $n_{divers}$ = 6) were included in the analysis. The mean age of the participants was *M* = 32.90 years (*SD* = 10.87), with an age range of 18 to 74 years (Figure 1).

**Figure 1**

*Age distribution of participants*

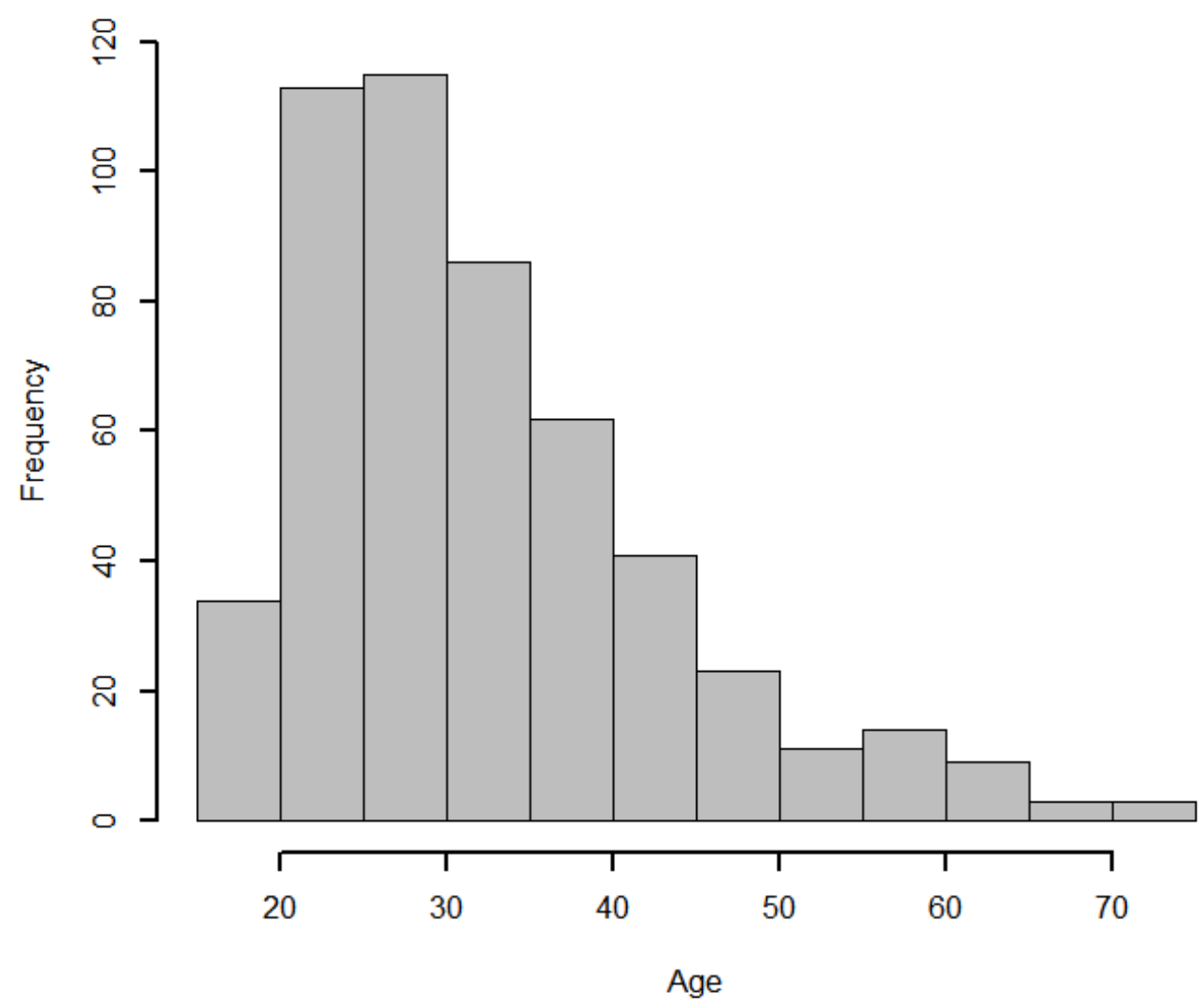

Regarding educational attainments, most participants stated they had either a Bachelor's degree or a German Abitur (equivalent to a high school diploma) (26.84% each). A smaller proportion reported holding a Master's degree (17.12%), followed by various other educational attainments (Table 3). Participants were also asked to indicate their professional or academic background concerning their current job. There were eleven categories to choose from, based on the defined occupational fields of the German Federal Employment Agency (2024), which were grouped into two main categories: technical and economic professions and social, creative, and service-oriented professions. Table 4 shows that most participants assigned themselves to Informatics/Data Science (20%) or Administration/Management (18%). Information on the participants' study status was obtained to further characterize the sample. At the time of the study, 38% of respondents were studying (*n* = 196), while 55% were not studying (*n* = 283). 7% did not provide information (*n* = 35). Current employment was also recorded. Most participants (41%) were employed full-time (Table 4). Information on student status and current employment status was voluntary. To assess participants' existing experience and use of AI, they were asked about their use and subjective prior knowledge of AI. AI use was assessed using two items that measured professional and personal use of AI on a five-point Likert scale (1 = "almost never", 5 = "daily") (Kanniainen et al., 2022; Pinski & Benlian, 2023). Subjective prior AI knowledge was measured on a scale from 1 ("poor") to 5 ("very good") (Hayo & Neuenkirch, 2014; Manika et al., 2021). Average AI use (*M* = 3.34, *SD* = 1.06) and prior AI knowledge (*M* = 3.11, *SD* = 0.91) were in the medium range.

**Table 3**

*Distribution of educational attainments of the participants in the sample*

| German educational attainment | Equivalent educational attainment in the USA | Frequency (n) | Frequency (%) |
|---|---|---|---|
| No qualification | - | 0 | 0 |
| Hauptschulabschluss* | High School Diploma (Low Academic Track) | 13 | 2.53 |
| Realschulabschluss* | High School Diploma (General Track) / Vocational Education | 70 | 13.62 |
| Fachhochschulreife* | High School Diploma (Vocational Track) / Community College | 29 | 5.64 |
| Abitur* | High School Diploma (College Prep) / AP / IB | 138 | 26.84 |
| Bachelor's Degree | - | 138 | 26.84 |
| Master's Degree | - | 88 | 17.12 |
| Diploma | - | 20 | 3.89 |
| Doctoral Degree | Doctorate (PhD) | 13 | 2.53 |
| Habilitation | Professorship / Post-doctoral qualification | 0 | 0 |
| Other qualification | - | 5 | 0.97 |

*Note.* *Hauptschulabschluss = Lower secondary school leaving certificate; *Realschulabschluss = Intermediate school leaving certificate; *Fachhochschulreife = Higher education entrance qualification (for universities of applied sciences); Abitur = General higher education entrance qualification (A-levels)

**Table 4**

*Distribution of professional or academic background and employment status in the sample*

| Demographics | | Frequency (n) | Frequency (%) |
|---|---|---|---|
| **Professional / academic background** | ***Technical and economic professions*** | | |
| | Administration / Management | 92 | 17.90 |
| | Craft / Technology | 15 | 2.92 |
| | Science / Research | 50 | 9.73 |
| | Informatics / Data Science | 105 | 20.62 |
| | Production / Manufacturing | 10 | 1.95 |
| | Agriculture / Construction / Environment | 5 | 0.97 |
| | ***Social, creative, and service-oriented professions*** | | |
| | Health / Care | 36 | 7.00 |
| | Social affairs / education | 43 | 8.37 |
| | Art / Culture / Media | 38 | 7.39 |
| | Services / Trade | 58 | 11.28 |
| | ***Other*** *(= none of the proposed categories fit)* | 62 | 12.06 |
| **Employment status** | Full-Time | 212 | 41.25 |
| | Part-Time | 121 | 23.54 |
| | Not in paid work (e.g., homemaker, retired, or disabled) | 26 | 5.06 |
| | Due to start a new job within the next month | 15 | 2.92 |
| | Unemployed / job-seeking | 54 | 10.51 |
| | Other *(= none of the proposed categories fit)* | 46 | 8.95 |
| | No Answer | 40 | 7.78 |

## 2.3 Measures and Validation

The following measurement methods were used to assess the reliability and validity of the measurement instrument.

***Reliability.*** Reliability is crucial to ensure that the measurement instrument consistently captures the construct of interest across different respondents and conditions (Nunnally & Bernstein, 1994). Several analyses have been used to assess the reliability of the AICOS. These include IRT-specific reliabilities such as expected a posteriori reliability (EAP), which measures the accuracy of individual ability estimates based on the IRT model, and empirical reliability, which indicates the accuracy of ability estimates in a given sample based on actual item responses, with values of .70 and above considered satisfactory (Raju et al., 2007). Conditional reliability was also calculated, which measures the precision of the ability estimate for a given ability level (Raju et al., 2007). Composite reliability (*CR*) and Cronbach's alpha ($\alpha$) were calculated to assess internal consistency. Values above .60 are acceptable in the early stages of research, and values above .70 are considered satisfactory (Hair et al., 2014).

***Structural validity.*** To evaluate the structural validity of the AICOS, a confirmatory factor analysis (CFA) was performed, with *RMSEA*, *SRMR*, *CFI*, and *TLI* as fit indices (Kline, 2023).

***Convergent validity.*** Convergent validity must be established to ensure that a measurement instrument accurately captures the intended construct and shows appropriate correlations with theoretically related constructs (Kline, 2023). To determine convergent validity, twelve items were selected from the previously validated measures of AI literacy by Hornberger et al. (2023) ($\alpha$ = .82) and Pinski et al. (2023) ($\alpha$ = .84), hereafter referred to as the *H&P-items* (Table A1). As the items of these measures are also part of the initial item pool of this study, the selection was 1) randomized and 2) ensured that the selected items are largely covered by other items in terms of content, so that no gaps in content arise. To determine convergent validity, a correlation between scores of the AICOS with other established measurement instruments of the same or theoretically related construct should exceed $r \geq .30$ (Cronbach & Meehl, 1955; Gregory, 2004).

***Discriminant validity.*** Discriminant validity must be established to demonstrate that a construct is distinct from related but conceptually different constructs (Kline, 2023). To this end, subjective AI literacy was measured using the short version of the Meta AI Literacy Scale by Koch et al. (2024) (10 items, $\alpha$ = .77). The scale comprises five subscales: "Use & Apply AI", "Know & Understand AI", "Detect AI", "Evaluate & Create AI", and "AI Ethics". Items are rated on an 11-point Likert scale (0 = not at all or hardly pronounced; 10 = very pronounced or almost perfectly pronounced). Studies show that subjective and objective knowledge are often uncorrelated or only weakly correlated — a pattern also observed for AI literacy (Carlson et al., 2009; Klerck & Sweeney, 2007; Ma & Chen, 2023; Moore & Healy, 2008; Pinski et al., 2024). Discriminant validity can be assumed if correlations between AICOS and unrelated constructs remain at or below $r = .20$ (Cohen, 1992; Lewis, 2003; Robinson, 2018).

***Criterion validity.*** Criterion validity is important because it demonstrates how well a measurement instrument correlates with relevant external criteria, thereby ensuring its practical applicability and predictive power (Cronbach & Meehl, 1955). To measure criterion validity, the scores of the AICOS were compared with the educational or professional background (concurrent validity) and the results of a follow-up test (predictive validity). In previous studies, individuals with technical backgrounds, such as engineering and computer science, demonstrated higher AI literacy compared to those with other educational backgrounds, including social sciences (Hornberger et al., 2023; Mansoor et al., 2024). Concurrent validity is assumed if participants with a computer science background perform better in the AICOS than people from other educational backgrounds. In addition, a follow-up quiz consisting of five practical MCQs requiring technical (especially IT) knowledge was developed (Table A2). This quiz was designed based on previous computer science and data science exams (e.g., https://www.statistik.rw.fau.de/klausuren/). Predictive validity is assumed if higher scores in the AICOS predict higher scores in the follow-up quiz and there is a correlation of $r \geq .40$ (Peers, 2006).

***Control questions.*** Four control questions (Table A3) were included in the study to ensure the quality of the data collected. They serve to identify inattentive participants, thus ensuring the reliability and validity of the results. The selection and construction of the control questions were based on the work of Barone et al. (2015) and Meade and Craig (2012).

### 2.4 Procedure

The study was conducted online using the *SoSci Survey* (Leiner, 2024). The median time to complete the survey was 62.78 minutes. After agreeing to the data collection and privacy policy, participants received an introduction explaining what was expected of them in this study. Subjective AI knowledge and frequency of AI use were then recorded. Next, they were given an instructional framework for the items of the AICOS (*n* = 107, items resulted from phase 2), explaining that only one answer per item was correct. In addition, participants were asked to complete the AICOS without using any aids (e.g., Internet searches for answers), which they consented to by checking an additional box. Consent was also obtained for JavaScript scripts in the survey to check the time of absence or tab changes during the survey to detect possible signs of cheating. Afterward, the participants answered the items in the AICOS presented in a randomized order. There was no time limit for completing the test, and participants were asked to complete it without interruption and in quiet surroundings. Participants then completed validation questionnaires (H&P items, MAILS, follow-up test) and provided demographic information (gender, age, educational background, professional/study-related field of activity) before being informed of the purpose of the study.

### 2.5 Item Response Theory Analysis

***IRT model selection.*** IRT models have been used to determine the characteristics of test items and the abilities of individuals on a common scale (Embretson & Reise, 2009). Well-known IRT models are the 1PL/Rasch model, the 2PL model, and the 3PL model. They differ in the number of item parameters estimated. In the 1PL model, only the parameter for item difficulty (*b*-parameter) is estimated. The 2PL model also estimates the item discrimination parameter (*a*-parameter), and the 3PL model includes an additional guessing parameter (*c*-parameter). Model fit in the IRT is assessed using the *M2* statistic, with a low *M2* value and a high or non-significant *p*-value indicating a good fit between model and data (Maydeu-Olivares & Joe, 2006). The fit indices *RMSEA* and *SRMR* ($\leq$ 0.05 for acceptable) and *CFI* and *TLI* ($\geq$ 0.90 for acceptable, $\geq$ 0.95 for very good) were also reported in the *M2* statistic (Brown, 2015; Maydeu-Olivares, 2013). Finally, the *MNSQ* statistic is considered, where the infit and outfit values should be in the range of 0.50-1.50 for an appropriate model fit (Ames & Penfield, 2015; Linacre, 2017) or, more strictly, in the range of 0.70-1.30 (Bond & Fox, 2013).

***Interpretation of IRT Parameters.*** There are guidelines for evaluating the psychometric quality of items for each of the IRT parameters (Table 5): For the *b*-parameter, which indicates the ability value at which there is a 50% probability of a correct response, values between -3 (very easy) and 3 (very difficult) are considered acceptable (Baker, 2001; Borges et al., 2017). The *a*-parameter describes how strongly an item discriminates between people of different ability levels. Values significantly below 0.35 or above 2.5 indicate overfitting or underfitting, respectively, and exclusion is recommended (Baker, 2001; Reeve & Fayers, 2005). The *c*-parameter indicates the probability of getting a correct answer by guessing. Items with *c*-parameters $\leq$ 0.35 are recommended for elimination as they may interfere with test information (Baker, 2001).

**Table 5**

*Guidelines for evaluating the item parameters in the item response analysis*

| Parameters | Interpretation/Quality (Baker, 2001) | Elimination guideline for this study |
|---|---|---|
| b (difficulty) | Acceptable if b = [-3; 3]<br>b ≤ -2.00 (very easy)<br>b = -2.00 ≤-1.00 (easy)<br>b = -1.00 ≤1.00 (moderately)<br>b = 1.00 ≤ 2.00 (difficult)<br>b > 2.00 (very difficult) | Items with values well above 3 or below -3 |
| a (discrimination) | a = 0 (no discrimination)<br>a = 0.01 - 0.34 (very low)<br>a = 0.35 - 0.64 (low)<br>a = 0.65 - 1.34 (moderate)<br>a = 1.35 - 1.69 (high)<br>a >1.70 (very high) | Clearly below 0.35 or above 2.5 |
| c (guessing) | c ≤ 0.35 | Items with values above 0.35 |

***IRT requirements.*** Before estimating IRT parameters, the assumptions of unidimensionality and local independence must be met (Embretson & Reise, 2000). **Unidimensionality** means that a single latent trait explains responses to all items. This is a core assumption in IRT, as it ensures that item responses are driven by one underlying construct, which is essential for valid parameter estimation and score interpretation (Embretson & Reise, 2000). It is tested using CFA and assessed with the following indices: $\chi^2/df < 2\text{-}3$, $RMSEA < 0.05$, $SRMR < 0.10$, $CFI \geq 0.90$, $TLI \geq 0.90$ (Brown, 2015). **Local independence** means that the response to an item depends only on the person's trait level and not on their responses to other items (Embretson & Reise, 2000). This assumption is essential in IRT to ensure that item parameters are unbiased and that inferences drawn from the model are valid (Yen, 1984). The Q3 statistic (Yen, 1984) was used to test local independence, with item pairs showing residual correlations above the threshold of 0.2, which is considered potentially problematic (Chen & Thissen, 1997).

# 3 RESULTS

## 3.1 Analysis of the Test Item Pool (107 items)

***Context effects.*** Given the long time required to complete the AICOS, whether fatigue affected the test results was examined. To minimize the effects of fatigue, items were randomized in presentation. The correlation between test duration and score was very low and insignificant ($r = -0.00$, $p = .946$). A Welch *t*-test ($t(509.93) = -0.12$, $p = 0.90$) showed no significant differences in scores between the "short" group ($M = 28.98$, $SD = 7.41$, $n = 257$) and the "long" group ($M = 28.98$, $SD = 7.41$, $n = 257$), which were calculated using a median split (*median* = 62.78 minutes), indicating that the randomization effectively minimized the potential impact of fatigue.

***Unidimensionality.*** Unidimensionality was assessed using a CFA with a unifactorial model. The *WLSMV* estimator (Weighted Least Squares Mean and Variance adjusted) was applied, as it provides reliable estimates for categorical data (Beauducel & Herzberg, 2006; Kılıç & Doğan, 2021). The results indicate a very good model fit ($\chi^2 = 5751.67$, $df = 5564$, $p = .039$, $\chi^2/df = 1.04$, $RMSEA = .008$, $SRMR = .082$, $CFI = .957$, $TLI = .956$). The high *CFI* and *TLI* values indicate a good global fit of the single-factor model, whereas the *SRMR* is somewhat elevated. This pattern suggests that residual correlations between certain item pairs are larger than expected under the single-factor model, pointing to potential local dependencies to which *CFI* and *TLI* are relatively insensitive (Brown, 2015; Schermelleh-Engel et al., 2003). Combined with inter-item correlations ranging from -.18 to .45 (cf. Supplementary Material 3), including some negative values, these patterns may reflect mis-specified items or multidimensionality (Brown, 2015; DeVellis, 2017). However, *SRMR* values should be interpreted with caution, as they are prone to bias in binary-coded data, such as in this study (Yu, 2002).

***Selection of IRT model.*** The IRT models were estimated using the *mirt* package (Chalmers, 2012) for *R* (R Core Team, 2013). The *L-BFGS-B* algorithm, a memory-efficient variant of the *BFGS* method for solving nonlinear optimization problems, was applied (Alcalá-Quintana & García-Pérez, 2013). All three classical IRT models (1PL, 2PL, and 3PL) were fitted, and the M2 statistic was computed. Model fit indices in Table 6 indicate that the 3PL model provides a better fit than the 1PL and 2PL models, as evidenced by the *M2* score, *RMSEA*, *CFI*, and *TLI*. The *p*-value for the 3PL model is the highest among the tested models, indicating superior fit (Maydeu-Olivares & Joe, 2006). The *SRMSR* is similar for the 3PL and 2PL models, while the 1PL model performs worse. Overall, the fit indices for the 2PL and 3PL models fall within the good to very good range (Brown, 2015; Maydeu-Olivares, 2013), with the 3PL model performing the best. The *MNSQ* statistic supports the 3PL model fit, as both Infit and Outfit values (Table A4) fall within conventional [0.50, 1.50] and stricter [0.70, 1.30] intervals (Bond & Fox, 2013; Linacre, 2017). Therefore, the 3PL model was selected for further analysis.

**Table 6**

*Fit indices of the item response models of the test item pool*

| Modell | *M2* | *df* | *p* | *RMSEA* | *SRMSR* | *CFI* | *TLI* |
|---|---|---|---|---|---|---|---|
| 1PL | 8587.46 | 5670 | < .001 | 0.032 | 0.068 | 0.816 | 0.816 |
| 2PL | 6150.39 | 5564 | < .001 | 0.014 | 0.044 | 0.963 | 0.962 |
| 3PL | 5653.28 | 5457 | .031 | 0.008 | 0.043 | 0.988 | 0.987 |

***Local independence.*** The Q3 statistic (Yen, 1984) based on the 3PL model assessed local independence for the IRT analysis. All residual correlations were below .20, supporting local independence, except for the correlation between the labeled Items EA13 and CA12 ($r$ = .21), which exceeded the critical threshold of .20 (Chen & Thissen, 1997). Given the higher number of EA items compared to CA items, EA13 was excluded to maintain local independence. The labels for the items can be found in Supplementary Material 3.

***Results of IRT analysis.*** The IRT analysis of the 107 items from the test item pool was based on the 3PL model, with difficulty parameters ($b$), discrimination coefficients ($a$), and guessing probabilities ($c$) calculated. Items with problematic parameters were considered for exclusion based on predefined criteria (see Section 2.5). Table 7 summarizes the relevant parameters, with detailed results in Table A5. Although the psychometric parameters of the items (AA21: $a = 3.28$, AA23: $b = 3.95$, DA05: $c = 0.39$, DA08: $a = 0.33$, UA21: $a = 2.84$, UA28: $c = 0.75$) do not meet the conservative exclusion criteria commonly found in the psychometric literature, we decided to retain them after a thorough content review. These items capture key aspects of AI competence not addressed by other items in the test, which we considered essential for a comprehensive assessment. Following this decision, 51 items were retained, and the IRT was rerun to reassess the model estimates for these items.

**Table 7**

*Descriptive summary of the IRT parameter values of the test item pool*

| Parameter | M | SD | Min | Max | Number of potential eliminations |
|---|---|---|---|---|---|
| Difficulty (b) | 0.49 | 6.28 | -11.26 | 53.57 | 13 |
| Discrimination (a) | 0.95 | 1.74 | -13.15 | 4.41 | 26 |
| Guessing (c) | 0.24 | 0.22 | 0.00 | 0.91 | 29 |

### *3.2* Analysis of the Final Item Pool (51 items)

***Unidimensionality.*** The assumption of unidimensionality was tested as outlined in Section 3.1. The fit indices indicate that the model exhibits an acceptable to very good fit with the data ($\chi^2 = 1287.14$, $df = 1224$, $p = .102$, $\chi^2/df = 1.05$, *RMSEA* = .010, *SRMR* = .068, *CFI* = .980, *TLI* = .979). The inter-item correlations ranged from 0.01 to 0.45 (see Supplementary Material 3) and were consistently positive, supporting the unidimensionality of the scale (Brown, 2015; DeVellis, 2017).

***Selection of IRT model.*** The selection of the IRT model followed the procedure outlined in Section 3.1. The 3PL model exhibited the lowest *M2* and *RMSEA* scores, the highest *CFI* and *TLI* values, and a non-significant $p$-value, indicating a good model fit (Maydeu-Olivares & Joe, 2006) (Table 8). The *SRMSR* is comparable between the 3PL and 2PL models. The Infit and Outfit scores fall within the recommended interval limits (Bond & Fox, 2013) (Table A6). Overall, the 3PL model demonstrates a very good fit and was selected for further analysis.

**Table 8**

*Fit indices of the item response models of the final item pool*

| Modell | *M2* | *df* | *p* | *RMSEA* | *SRMSR* | *CFI* | *TLI* |
|---|---|---|---|---|---|---|---|
| 1PL | 1703.39 | 1274 | < .001 | 0.026 | 0.063 | 0.931 | 0.931 |
| 2PL | 1315.33 | 1224 | .035 | 0.012 | 0.041 | 0.985 | 0.985 |
| 3PL | 1212.40 | 1173 | .207 | 0.008 | 0.041 | 0.993 | 0.993 |

***Local independence.*** The local independence was calculated as described in Section 3.1. The Q3 statistics, based on the 3PL model, show that all residual correlations were below $r = .20$, confirming local independence.

***Results of IRT analysis.*** In the IRT analysis, the parameters of the final item pool were evaluated for their psychometric properties according to the criteria set out in Section 3.2. Table 9 summarizes the relevant parameters of all items. Table 10 provides detailed results for the specific parameters of each item. Two items slightly exceed the established thresholds: CA04 with a *c*-parameter of 0.34 and AA23 with a *b*-parameter of 3.34. Since these exceedances are marginal and higher *b*-values can better differentiate the upper end of the ability spectrum, it was decided to retain these items. As a result, the 51 items were retained in the final pool, with the full list available in Supplementary Material 4.

**Table 9**

*Descriptive summary of the IRT parameter values of the final item pool*

| Parameter | M | SD | Min | Max | Number of potential eliminations |
|---|---|---|---|---|---|
| Difficulty (b) | 0.00 | 1.34 | -2.84 | 3.34 | 1 |
| Discrimination (a) | 1.09 | 0.49 | 0.34 | 2.20 | 1 |
| Guessing (c) | 0.13 | 0.13 | 0.00 | 0.35 | 0 |

**Table 10**

*Item parameters and source of the 51 items in the final item pool*

| | **IRT-parameter** | | | **Source** |
|---|---|---|---|---|
| | a | b | g | **adapted from** |
| **AA02** | 0.63 | -2.05 | 0.00 | Zhang et al. (2024) |
| **AA04** | 1.05 | 1.00 | 0.26 | Hornberger et al. (2023) |
| **AA06** | 1.81 | 2.10 | 0.19 | Kong et al. (2024) |
| **AA08** | 1.32 | 1.08 | 0.07 | Zhang et al. (2024) |
| **AA09** | 0.64 | -1.19 | 0.01 | MOTIV |
| **AA19** | 1.14 | -0.14 | 0.27 | Yau et al. (2022) |
| **AA20** | 0.79 | 0.29 | 0.00 | Hornberger et al. (2023) |
| **AA21** | 1.78 | 2.30 | 0.21 | MOTIV |
| **AA22** | 1.18 | 0.93 | 0.33 | Chiu et al. (2024) |
| **AA23** | 0.81 | 3.34 | 0.23 | Chiu et al. (2024) |
| **CA02** | 1.45 | 0.90 | 0.26 | Kong et al. (2024) |
| **CA03** | 0.53 | -0.90 | 0.10 | KI Campus |
| **CA04** | 0.34 | -0.34 | 0.00 | Self-created |
| **CA07** | 1.37 | -0.11 | 0.34 | Rodríguez-García et al. (2021) |
| **CA10** | 0.68 | 0.81 | 0.00 | Chiu et al. (2024) |
| **CA11** | 1.79 | 0.36 | 0.13 | Hornberger et al. (2023) |
| **CA12** | 1.37 | -1.40 | 0.00 | Elements of AI |
| **DA01** | 0.58 | -2.17 | 0.00 | Hornberger et al. (2023) |
| **DA02** | 0.74 | 1.67 | 0.33 | Hornberger et al. (2023) |
| **DA03** | 1.30 | 0.50 | 0.22 | Pinski et al. (2024) |
| **DA04** | 0.61 | -1.42 | 0.00 | Hornberger et al. (2023) |
| **DA05** | 1.13 | -0.09 | 0.32 | KI Campus |
| **DA08** | 0.36 | -0.76 | 0.00 | Self-created |
| **DA09** | 0.99 | -1.30 | 0.00 | Pinski et al. (2024) |
| **DA10** | 0.89 | 1.60 | 0.08 | MOTIV |
| **EA01** | 0.74 | -0.74 | 0.08 | Hornberger et al. (2023) |
| **EA02** | 1.95 | -0.14 | 0.26 | Hornberger et al. (2023) |
| **EA05** | 1.40 | -1.28 | 0.00 | MOTIV |
| **EA09** | 1.84 | -0.46 | 0.28 | KI Campus |
| **EA11** | 1.29 | -0.66 | 0.00 | KI Campus |
| **EA15** | 0.85 | 0.16 | 0.21 | Chiu et al. (2024) |
| **GA02** | 1.42 | -1.63 | 0.00 | Self-created / Annapureddy et al. (2024) |
| **GA04** | 0.80 | 1.01 | 0.14 | Self-created / Annapureddy et al. (2024) |
| **GA08** | 0.98 | 1.06 | 0.14 | Self-created / Annapureddy et al. (2024) |
| **GA10** | 1.00 | -1.45 | 0.00 | Self-created / Annapureddy et al. (2024) |
| **GA11** | 0.41 | -1.29 | 0.00 | Self-created / Annapureddy et al. (2024) |
| **GA13** | 0.75 | -1.73 | 0.17 | Self-created / Annapureddy et al. (2024) |
| **GA16** | 0.96 | -1.01 | 0.13 | Self-created / Annapureddy et al. (2024) |
| **GA17** | 1.63 | 0.11 | 0.34 | Self-created / Annapureddy et al. (2024) |
| **GA19** | 0.66 | -1.00 | 0.00 | Self-created / Annapureddy et al. (2024) |
| **UA01** | 1.36 | 2.67 | 0.20 | Ding et al. (2024) |
| **UA05** | 1.01 | 1.54 | 0.26 | MOTIV |
| **UA09** | 1.54 | 1.20 | 0.26 | Hornberger et al. (2023) |
| **UA10** | 2.16 | 1.13 | 0.28 | Hornberger et al. (2023) |
| **UA11** | 0.42 | -1.05 | 0.01 | Hornberger et al. (2023) |
| **UA18** | 0.37 | -0.25 | 0.00 | MOTIV |
| **UA19** | 0.60 | 0.23 | 0.00 | Yau et al. (2022) |
| **UA21** | 2.20 | 1.32 | 0.35 | Self-created |
| **UA26** | 1.63 | -0.65 | 0.00 | MOTIV |
| **UA27** | 1.28 | 0.88 | 0.26 | Kong et al. (2024) |
| **UA28** | 0.95 | -2.84 | 0.00 | Self-created |

***Model Visualization.*** Figure 2 presents the Item Characteristic Curves (ICCs), illustrating the relationship between a person's ability ($\theta$) and the probability of correctly answering an item. All items exhibit an S-shaped curve, consistent with the expected increase in response probability as ability rises. Figure 3 presents a Wright Map illustrating the relationship between person abilities and item difficulties. Like the ICCs, the items are evenly distributed across the ability range ($M$ = 0.00, $SD$ = 0.93), with most ability levels covered by items of varying difficulty. Item AA23 represents the upper ability range ($Logit \approx +3$), while item UA28 covers the lower range ($Logit \approx -3$). The test spans the recommended ability range of -2 to +2 (Baker, 2001; De Ayala, 2013), although gaps exist in the extreme ability ranges ($Logit \approx$ +2 to +3 and -2 to -3), limiting measurement precision in these areas. Figure 4 also shows the Test Information Curve (TIF), confirming high accuracy in the middle ability range ($\theta$ = 0.47) and reduced accuracy at the extremes. The highest point on the test information curve is 7.21, with an average TIF of 5.32.

**Figure 2**

*Item Characteristic Curves (ICCs) showing the relationship between ability (θ) and the probability of correct responses across all items*

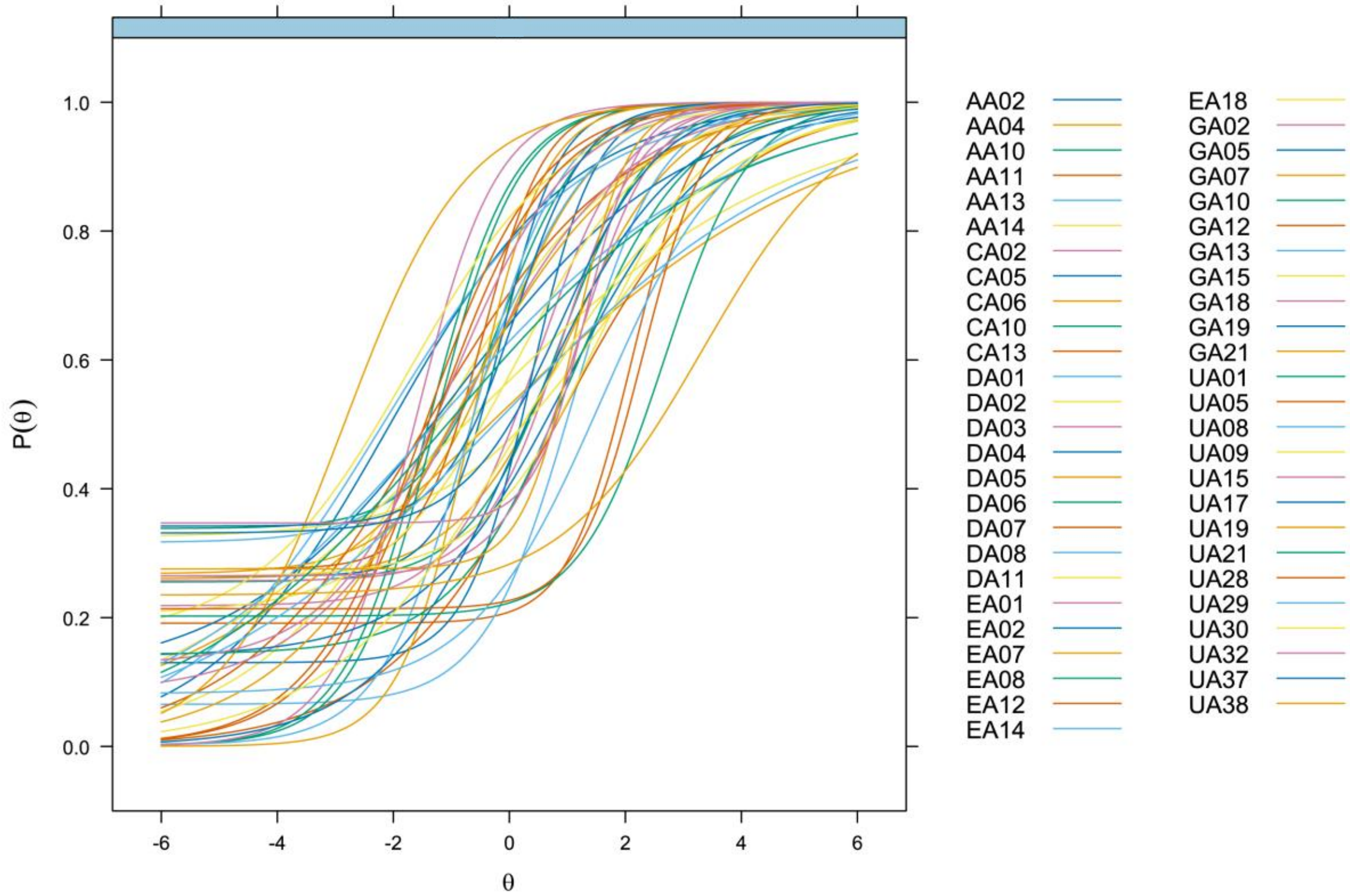

**Figure 3**

*Wright Map displaying the distribution of person abilities and item difficulties*

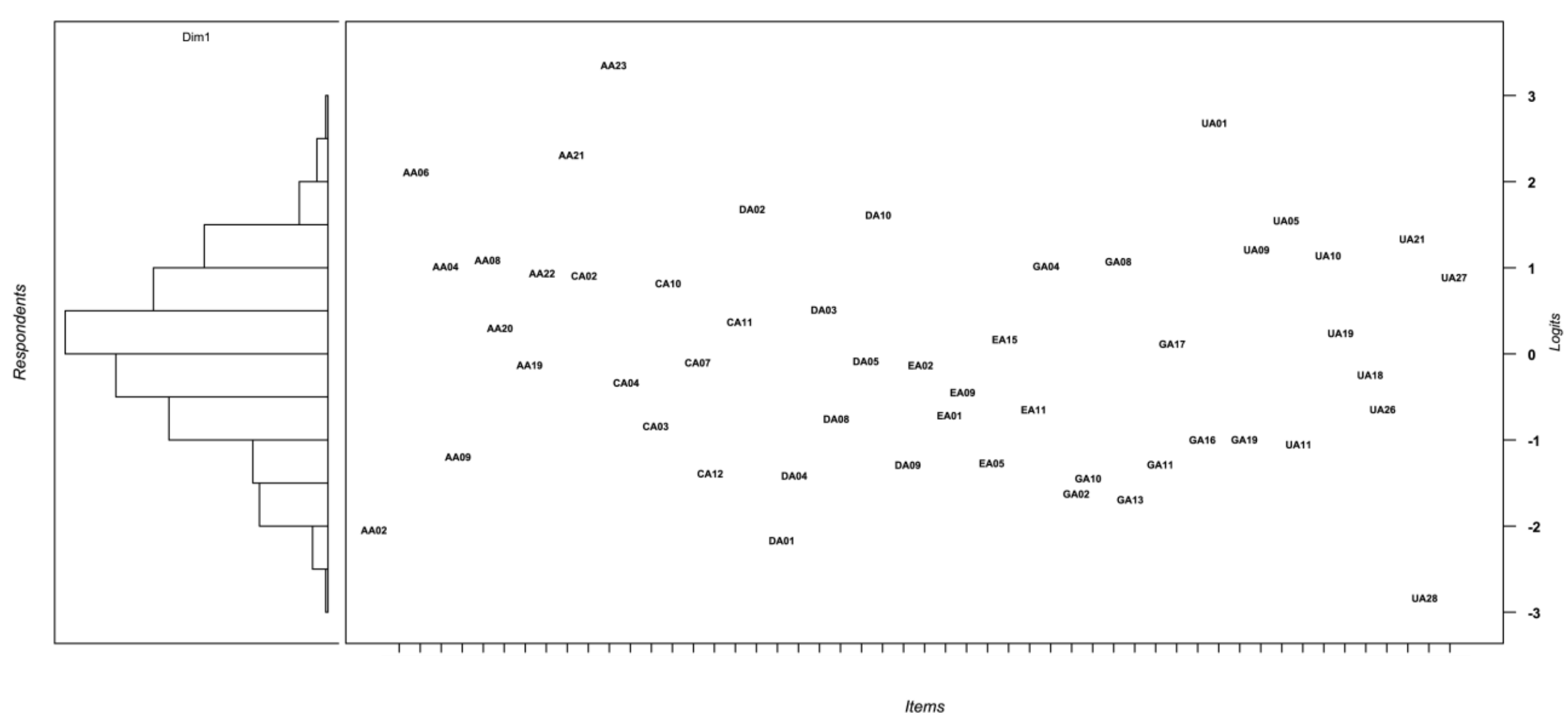

**Figure 4**

*Test Information Curve (TIF) illustrating test accuracy*

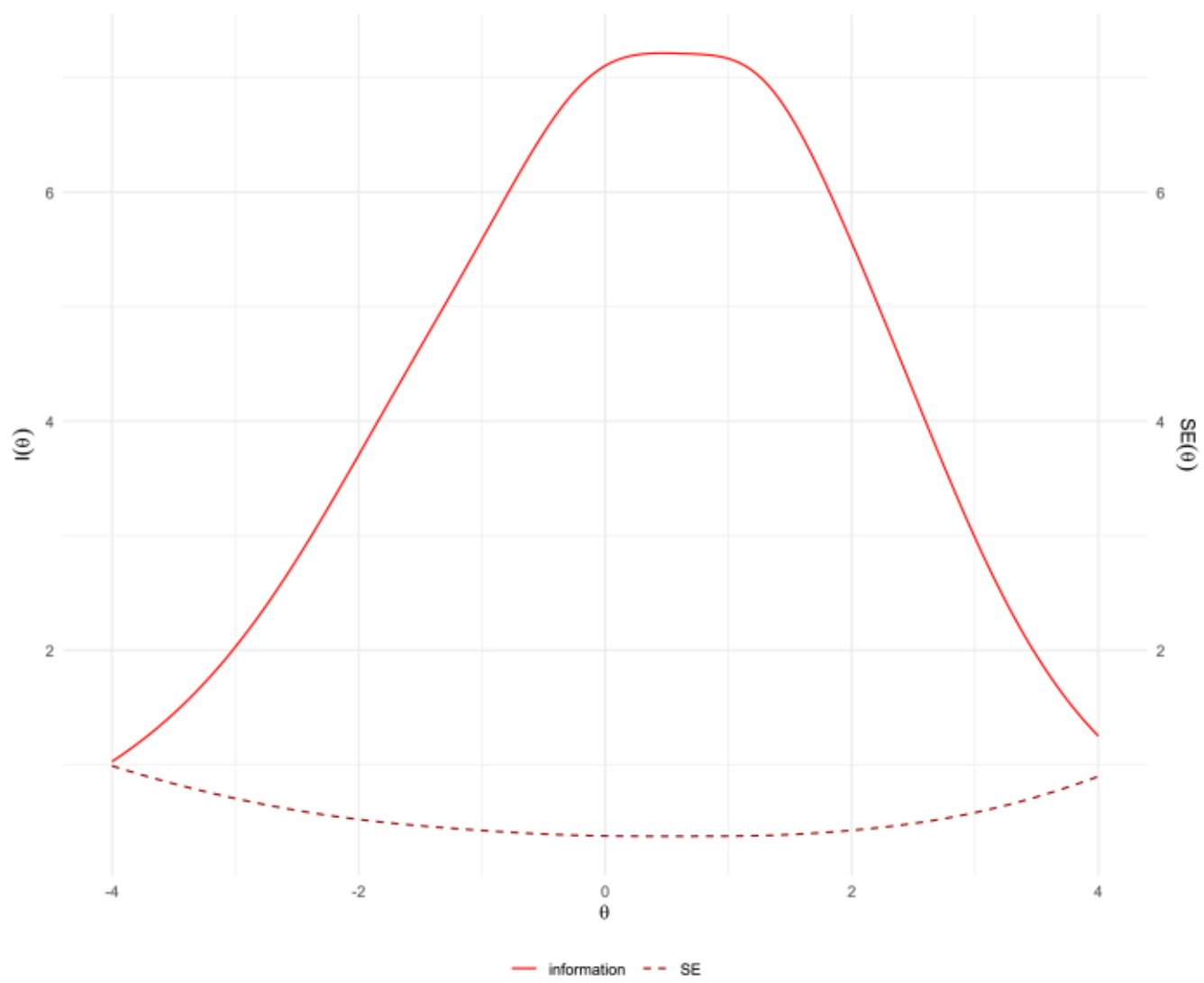

### 3.3 Reliability

Multiple reliability indices were examined: Cronbach's alpha ($\alpha$ = .83) and Composite Reliability (*CR* = .90) indicate good internal consistency of the AICOS. The reliability of the EAP ability estimates (*r* = .83) and the empirical reliability (*r* = .86), which evaluates the consistency of test performance based on observed data, further support the test's reliability. Conditional reliability, which accounts for the precision of ability estimates across different ability levels, confirms an overall high measurement precision (*r* = .91). The reliability values range from .77 (for $\theta = 3$) to .93 (for $\theta = -0.7$), indicating acceptable to very high precision across ability/theta levels (Table A7).

### 3.4 Validity

The CFA conducted on the final item pool indicated good model fit (cf. Section 3.2), supporting the structural validity of the AICOS (Brown, 2015; Kline, 2023). The significant correlation between AICOS and H&P test scores (both calculated as the sum of correctly solved items; $r = .58$, $p < .001$) provides evidence for *convergent validity* (Cronbach & Meehl, 1955; Gregory, 2004). Additionally, the low and non-significant correlation with the MAILS, which measures subjective AI Literacy ($r = .04$, $p = .424$), supports *discriminant validity* (Cohen, 1992; Lewis, 2003). Both concurrent and predictive validity were examined to assess criterion validity. The significantly higher competence scores of people with a computer science background ($M = 0.33$, $SD = 0.94$, $n = 105$) compared to people without a computer science background ($M = -0.09$, $SD = 0.91$, $n = 409$) provide evidence for the *concurrent validity* of the AICOS ($F(1, 512) = 17.45$, $p < .001$, $\eta^2 = 0.05$). Furthermore, scores of AICOS showed a significant correlation with the results of the follow-up quiz ($r = .42$, $p < .001$), supporting the *test's predictive validity* (Peers, 2006).

### 3.5 Outlook On Demographic Effects

This section provides an outlook on potential demographic differences in AI literacy. However, given the early stage of scale development, these findings should be interpreted cautiously and further explored in future research. Correlations with AI literacy scores were analyzed using point-biserial correlation ($r_{pb}$) for nominal data, *Spearman's Rho* for ordinal data, and Pearson's correlation (*r*) for interval-scaled data. Results showed a negative correlation with gender (0 = male; 1 = female, $r_{pb} = -.10$, $p = .042$) and a positive correlation with educational level (*Spearman's Rho* = .31, $p < .001$). No significant correlations were found for age ($r = .01$, $p = .777$) or professional/academic background (*Spearman's Rho* = -

.02, $p = .652$). To further examine these correlations, group differences were analyzed using the nonparametric Kruskal-Wallis test with pairwise Dwass-Steel-Critchlow-Fligner comparisons to account for often unequal sample sizes.

***Gender.*** Significant differences in AICOS test scores were found between male ($n = 273$, $M = 29.66$, $SD = 7.45$) and female participants ($n = 235$, $M = 28.10$, $SD = 7.78$; $\chi^2(1) = 4.14$, $p = .042$, $\varepsilon^2 = 0.01$), with males scoring significantly higher ($W = -2.88$, $p = .042$).

***Educational attainment.*** Comparison of AICOS test scores revealed a significant difference between individuals with an academic degree ($n = 264$, $M = 31.00$, $SD = 7.54$) and those without ($n = 250$, $M = 26.93$, $SD = 7.21$) ($\chi^2(1) = 36.19$, $p < .001$, $\varepsilon^2 = 0.07$. Participants with an academic degree scored significantly higher ($W = 8.51$, $p < .001$). Significant differences were also found across specific German educational attainments (for English equivalents, see Table 3) ($\chi^2(8) = 58.18$, $p < .001$, $\varepsilon^2 = 0.11$). Compared to individuals with a *Realschulsabschluss* ($M = 23.99$, $SD = 7.69$), those with an *Abitur* ($M = 28.76$, $SD = 6.47$, $W = 5.93$, $p < .001$), Bachelor's degree ($M = 29.93$, $SD = 7.57$, $W = 6.94$, $p < .001$), Master's degree ($M = 32.64$, $SD = 7.67$, $W = 8.69$, $p < .001$), Diploma ($M = 30.90$, $SD = 6.23$, $W = 4.71$, $p = .025$), and Doctoral degree ($M = 31.85$, $SD = 6.78$, $W = 4.65$, $p = .028$) achieved significantly higher scores. Higher education levels were associated with greater overall AI literacy, with Master's degree holders scoring higher than individuals with a *Fachhochschulreife* ($M = 25.75$, $SD = 7.01$, $W = 5.62$, $p = .002$) or *Abitur* ($W = 5.74$, $p = .002$).

***Age.*** To complement the correlation analysis, differences in AICOS test scores across age groups were examined: 18-29 ($n = 230$, $M = 28.60$, $SD = 7.80$), 30-39 ($n = 146$, $M = 30.40$, $SD = 7.11$), 40-49 ($n = 66$, $M = 28.90$, $SD = 7.56$), and 50-74 ($n = 42$, $M = 28.10$, $SD = 7.90$). The age groups 18-29, 30-39, and 40-49 reflect commonly used decade-based categorizations in empirical research. Due to small sample sizes in the older age ranges, these were combined into a broader group (50-74) for pragmatic reasons. No significant differences were found between the groups ($\chi^2(3) = 5.24$, $p = .154$, $\varepsilon^2 = 0.00$). Similarly, when using an alternative grouping that combines all individuals aged 40 and above into a group ($n = 108$, $M = 28.60$, $SD = 7.66$), no significant differences emerged ($\chi^2(2) = 5.09$, $p = .079$, $\varepsilon^2 = 0.01$).

***Professional / academic background.*** Significant differences were found for the major occupational categories ($\chi^2(1) = 15.96$, $p < .001$, $\varepsilon^2 = 0.04$) and the individual professional/academic backgrounds ($\chi^2(10) = 42.91$, $p < .001$, $\varepsilon^2 = 0.08$). Participants in technical and economic professions ($n = 277$, $M = 30.25$, $SD = 7.91$) scored higher than those in social, creative, and service-oriented professions ($n = 175$, $M = 27.50$, $SD = 7.19$, $W = 5.65$, $p < .001$). Individuals in Health/Care ($n = 36$, $M = 26.83$, $SD = 6.83$), Social Affairs/Education ($n = 43$, $M = 27.23$, $SD = 6.98$), and Service/Trade ($n = 58$, $M = 27.17$, $SD = 8.21$) scored significantly lower in AI literacy than those in Science/Research ($n = 50$, $M = 32.42$, $SD = 8.01$) and Informatics/Data Science ($n = 105$, $M = 31.74$, $SD = 7.64$). Post-hoc tests show that the differences between (1) Health/Care vs. Science/Research ($W = 5.03$, $p = .016$) and Informatics/Data Science ($W = 4.99$, $p = .018$), (2) Social Affairs/Education vs. Science/Research ($W = 4.86$, $p = .025$) and Informatics/Data Science ($W = 4.92$, $p = .022$), and (3) Service/Trade vs. Science/Research ($W = 4.70$, $p = .036$) and Informatics/Data Science ($W = 4.85$, $p = .026$) are significant.

### 3.6 Verification of the Six-Factor Structure

The AICOS was developed based on the AI-related sub-competencies outlined by Carolus, Koch, et al. (2023) and Annapureddy et al. (2024), with items allocated to six dimensions: Apply AI, Create AI, Detect AI, Ethics AI, Generative AI, and Understanding AI. A CFA confirmed a six-factor structure, with fit indices indicating good model fit ($\chi^2 = 1256.96$, $df = 1209$, $p = .164$, $\chi^2/df = 1.04$, $RMSEA = .009$, $SRMR = .067$, $CFI = .985$, $TLI = .984$). A robust $\chi^2$ difference test with the Satorra-Bentler correction (Satorra, 2000) was used to compare the one-factor model (cf. Section 3.2) with the six-factor model. The test showed a significant difference ($\Delta\chi^2 = 36.58$, $\Delta df = 15$, $p = .001$), with the six-factor model fitting slightly better. Standardized factor loadings ranged from $\beta = .11$ (UA01) to $\beta = .72$ (UA26). All items were significant at the 5% level except for UA01 ($p = .094$) and AA23 ($p = .057$), which showed marginal significance at the 10% level. Internal consistencies assessed via composite reliability ranged from .60 to .73, with values above .60 considered acceptable, particularly in the early stages of scale development (Hair et al., 2014). Table 11 provides an overview of the standardized factor loadings and *CR* values. To ensure the sub-competencies of AI literacy are sufficiently distinct, the heterotrait-monotrait ratio (HTMT) was calculated to assess the similarity between latent variables (Henseler et al., 2015). All values were below the .90 threshold, indicating sufficient distinctness of the constructs (Franke & Sarstedt, 2019). Results for the HTMT are shown in Table 12, with intercorrelations between subscales in Table 13. Overall, the results suggest a potential six-factor structure, but limited factor loadings, moderate subscale reliabilities, and only marginal model fit improvement support the one-factor solution as a methodologically justifiable approximation.

**Table 11**

*Overview of scale reliabilities and standardized item factor loadings*

| Subscale/Item | β | SE | z | p | Subscale/Item | β | SE | z | p |
|---|---|---|---|---|---|---|---|---|---|
| **Apply AI (*CR* = .60)** | | | | | **Create AI (*CR* = .64)** | | | | |
| AA02 | 0.39 | 0.07 | 5.74 | < .001 | CA02 | 0.40 | 0.06 | 6.97 | < .001 |
| AA04 | 0.36 | 0.06 | 5.60 | < .001 | CA03 | 0.30 | 0.06 | 4.93 | < .001 |
| AA06 | 0.25 | 0.07 | 3.46 | .001 | CA04 | 0.22 | 0.06 | 3.68 | < .001 |
| AA08 | 0.55 | 0.06 | 9.22 | < .001 | CA07 | 0.49 | 0.06 | 8.71 | < .001 |
| AA09 | 0.42 | 0.06 | 6.79 | < .001 | CA10 | 0.40 | 0.06 | 7.12 | < .001 |
| AA22 | 0.36 | 0.06 | 5.83 | < .001 | CA11 | 0.63 | 0.05 | 12.61 | < .001 |
| AA19 | 0.50 | 0.06 | 8.53 | < .001 | CA12 | 0.65 | 0.06 | 11.37 | < .001 |
| AA20 | 0.48 | 0.06 | 8.51 | < .001 | | | | | |
| AA23 | 0.13 | 0.07 | 1.91 | .057 | | | | | |
| AA21 | 0.17 | 0.07 | 2.42 | .016 | | | | | |
| **Detect AI (*CR* = .62)** | | | | | **Ethics AI (*CR* = .73)** | | | | |
| DA01 | 0.38 | 0.07 | 5.47 | < .001 | EA01 | 0.41 | 0.06 | 7.11 | < .001 |
| DA02 | 0.22 | 0.07 | 3.39 | .001 | EA02 | 0.62 | 0.05 | 12.82 | < .001 |
| DA03 | 0.51 | 0.06 | 8.68 | < .001 | EA05 | 0.66 | 0.05 | 12.30 | < .001 |
| DA04 | 0.40 | 0.07 | 6.21 | < .001 | EA09 | 0.63 | 0.05 | 12.75 | < .001 |
| DA05 | 0.49 | 0.06 | 8.11 | < .001 | EA11 | 0.63 | 0.05 | 13.18 | < .001 |
| DA08 | 0.26 | 0.07 | 3.90 | < .001 | EA15 | 0.38 | 0.06 | 6.64 | < .001 |
| DA09 | 0.61 | 0.06 | 9.69 | < .001 | | | | | |
| DA10 | 0.42 | 0.07 | 6.45 | < .001 | | | | | |
| **Generative AI (*CR* = .69)** | | | | | **Understanding AI (*CR* = .60)** | | | | |
| GA02 | 0.67 | 0.06 | 11.42 | <.001 | UA01 | 0.11 | 0.07 | 1.68 | .094 |
| GA04 | 0.35 | 0.06 | 5.99 | < .001 | UA05 | 0.28 | 0.06 | 4.66 | < .001 |
| GA08 | 0.39 | 0.06 | 6.50 | < .001 | UA09 | 0.34 | 0.06 | 5.69 | <.001 |
| GA10 | 0.55 | 0.06 | 9.64 | < .001 | UA10 | 0.36 | 0.06 | 5.97 | < .001 |
| GA11 | 0.27 | 0.06 | 4.34 | < .001 | UA11 | 0.27 | 0.06 | 4.47 | < .001 |
| GA13 | 0.38 | 0.06 | 5.91 | < .001 | UA18 | 0.24 | 0.06 | 4.03 | <.001 |
| GA16 | 0.48 | 0.06 | 8.57 | < .001 | UA19 | 0.37 | 0.06 | 6.62 | < .001 |
| GA17 | 0.50 | 0.06 | 9.13 | < .001 | UA21 | 0.26 | 0.06 | 4.16 | <.001 |
| GA19 | 0.39 | 0.06 | 6.72 | < .001 | UA26 | 0.72 | 0.05 | 15.58 | < .001 |
| | | | | | UA27 | 0.39 | 0.06 | 6.91 | < .001 |
| | | | | | UA28 | 0.46 | 0.08 | 5.92 | < .001 |

**Table 12**

*Heterotrait-Monotrait (HTMT) Ratios Between AI-Related Sub-Competencies*

| | AA | CA | DA | EA | GA | UA |
|---|---|---|---|---|---|---|
| Apply AI (AA) | 1.00 | | | | | |
| Crate AI (CA) | .743 | 1.00 | | | | |
| Detect AI (DA) | .596 | .724 | 1.00 | | | |
| AI Ethics (EA) | .776 | .867 | .826 | 1.00 | | |
| Generative AI (GA) | .833 | .875 | .702 | .880 | 1.00 | |
| Understanding AI (UA) | .798 | .884 | .753 | .714 | .859 | 1.00 |

**Table 13**

*Intercorrelations between the subscales of the AICOS*

| | AA | CA | DA | EA | GA | UA |
|---|---|---|---|---|---|---|
| Apply AI (AA) | | | | | | |
| Crate AI (CA) | .40*** | | | | | |
| Detect AI (DA) | .27*** | .38*** | | | | |
| AI Ethics (EA) | .41*** | .48*** | .43*** | | | |
| Generative AI (GA) | .42*** | .45*** | .39*** | .54*** | | |
| Understanding AI (UA) | .38*** | .44*** | .40*** | .44*** | .41*** | |

*Note*. *** = $p < .001$; $r < 0.30$ = small correlation; $r \geq 0.30$ and < 0.50 = medium correlation; $r \geq 0.50$ and < 0.70 = large correlation (Cohen, 1988).

### 3.7 Proposal for a Shortened Version of the AICOS

The AICOS consists of 51 items, which leads to a relatively long processing time (*median* = 21.71 minutes) and limits the instrument's economic efficiency. Therefore, a proposal for a potential short version of the AICOS is presented as a basis for future research projects. The construction of the shortened version followed the methodology of Koch et al. (2024), who selected items with the highest factor loadings for each subfactor and verified the factor structure using CFA. This approach ensured that the shortened version (SV) retained the theoretical structure of the scale while keeping the most informative items for each subfactor. Following this, an IRT analysis was conducted to assess item quality and ensure a balanced difficulty distribution in the SV, resulting in test information comparable to that of the long version (LV). In addition, it was considered that at least two to three items per construct are necessary to conduct a CFA to adequately capture the depth and complexity of the constructs (Bollen, 1989; Marsh et al., 1998). Preliminary analyses of the reliability and validity of the short version are also conducted to provide an initial assessment of its accuracy and quality. However, the need for validation with independent samples is emphasized to minimize potential bias. As a result, four potential short versions of the AICOS-SV were identified, with the one presented below being the most convincing in terms of reliability, validity, and test precision similarity. The other versions can be found in Supplementary Material 5.

The SV is based on a unidimensional model, which fits the data well ($\chi^2 = 151.07$, $df = 135$, $p = .164$, $\chi^2/df = 1.12$, $RMSEA = .015$, $SRMR = .063$, $CFI = .987$, $TLI = .985$) and showed overall acceptable reliability ($CR = .84$). For comparison, a hypothesized six-factor structure was also tested. Although it demonstrated acceptable fit ($\chi^2 = 131.08$, $df = 120$, $p = .231$, $\chi^2/df = 1.09$, $RMSEA = .013$, $SRMR = .059$, $CFI = .991$, $TLI = .988$), subscale reliabilities were suboptimal (CR: AA = .49, CA = .55, DA = .51, EA = .56, GA = .44, UA = .39) and the six-factor model did not significantly improve fit ($\Delta\chi^2 = 17.78$, $\Delta df = 15$, $p = .275$). Therefore, these results suggest that the unidimensional solution provides a more reliable and parsimonious representation for the SV. Comparison with the LV shows that factor loadings in the SV are in a similar range (LV: 0.11 – 0.70; SV: 0.24 – 0.71) and highly correlated between versions ($r = .99$). Regarding IRT parameters, the SV shows slightly lower average test information compared with the LV, as expected due to fewer items (LV = 5.32, SV = 3.82). The peak of the SV's test information curve is slightly shifted toward lower ability levels ($LV_\theta = 0.47$, $SV_\theta = 0.00$), indicating that the SV is somewhat easier than the LV. Descriptive comparisons further show that the SV retains similar average discrimination (LV a = 1.08, SV a = 1.19) and average difficulty (LV b = 0.00, SV b = -0.16) to the LV. Preliminary analyses indicate that the SV meets the required criteria for convergent validity (H&P-items: $r$ =

.58, $p < .001$) and discriminant validity (MAILS: $r = .02$, $p = .586$) while falling marginally below the threshold for criterion validity (quiz: $r = .38$, $p < .001$) (Cohen, 1992; Cronbach & Meehl, 1955; Gregory, 2004; Lewis, 2003; Robinson, 2018).

**Table 13**

*Items, standardized factor loadings, and item difficulties of the AICOS-SV*

| Items | β | SE | z | p | Discrimination (a) | Difficulty (b) |
|---|---|---|---|---|---|---|
| AA04 | 0.35 | 0.06 | 5.69 | < .001 | 1.35 | 1.06 → Difficulty |
| AA08 | 0.49 | 0.06 | 8.52 | < .001 | 0.98 | 1.02 → Difficulty |
| AA19 | 0.40 | 0.06 | 6.69 | < .001 | 1.29 | 0.22 → Moderately |
| CA02 | 0.36 | 0.06 | 5.91 | < .001 | 1.21 | 0.99 → Moderately |
| CA07 | 0.47 | 0.06 | 7.90 | < .001 | 1.06 | -0.47 → Moderately |
| CA11 | 0.63 | 0.05 | 12.43 | < .001 | 1.73 | 0.30 → Moderately |
| DA03 | 0.48 | 0.06 | 8.50 | < .001 | 1.02 | 0.08 → Moderately |
| DA05 | 0.46 | 0.06 | 8.02 | < .001 | 1.33 | -0.07 → Moderately |
| DA09 | 0.51 | 0.06 | 8.25 | < .001 | 1.01 | -1.29 → Easy |
| EA09 | 0.59 | 0.05 | 10.98 | < .001 | 1.4 | -0.85 → Moderately |
| EA11 | 0.63 | 0.05 | 12.40 | < .001 | 1.32 | -0.66 → Moderately |
| EA15 | 0.37 | 0.06 | 6.23 | < .001 | 0.97 | 0.40 → Moderately |
| GA02 | 0.64 | 0.06 | 10.64 | < .001 | 1.51 | -1.57 → Easy |
| GA04 | 0.34 | 0.06 | 5.64 | < .001 | 1.18 | 1.16 → Difficulty |
| GA08 | 0.41 | 0.06 | 6.78 | < .001 | 0.82 | 0.85 → Difficulty |
| UA11 | 0.24 | 0.06 | 3.93 | < .001 | 0.44 | -0.91 → Moderately |
| UA26 | 0.71 | 0.05 | 14.74 | < .001 | 1.73 | -0.65 → Moderately |
| UA28 | 0.44 | 0.08 | 5.68 | < .001 | 1.12 | 0.90 → Moderately |

**Figure 5**

*Test information curve of AICOS short version*

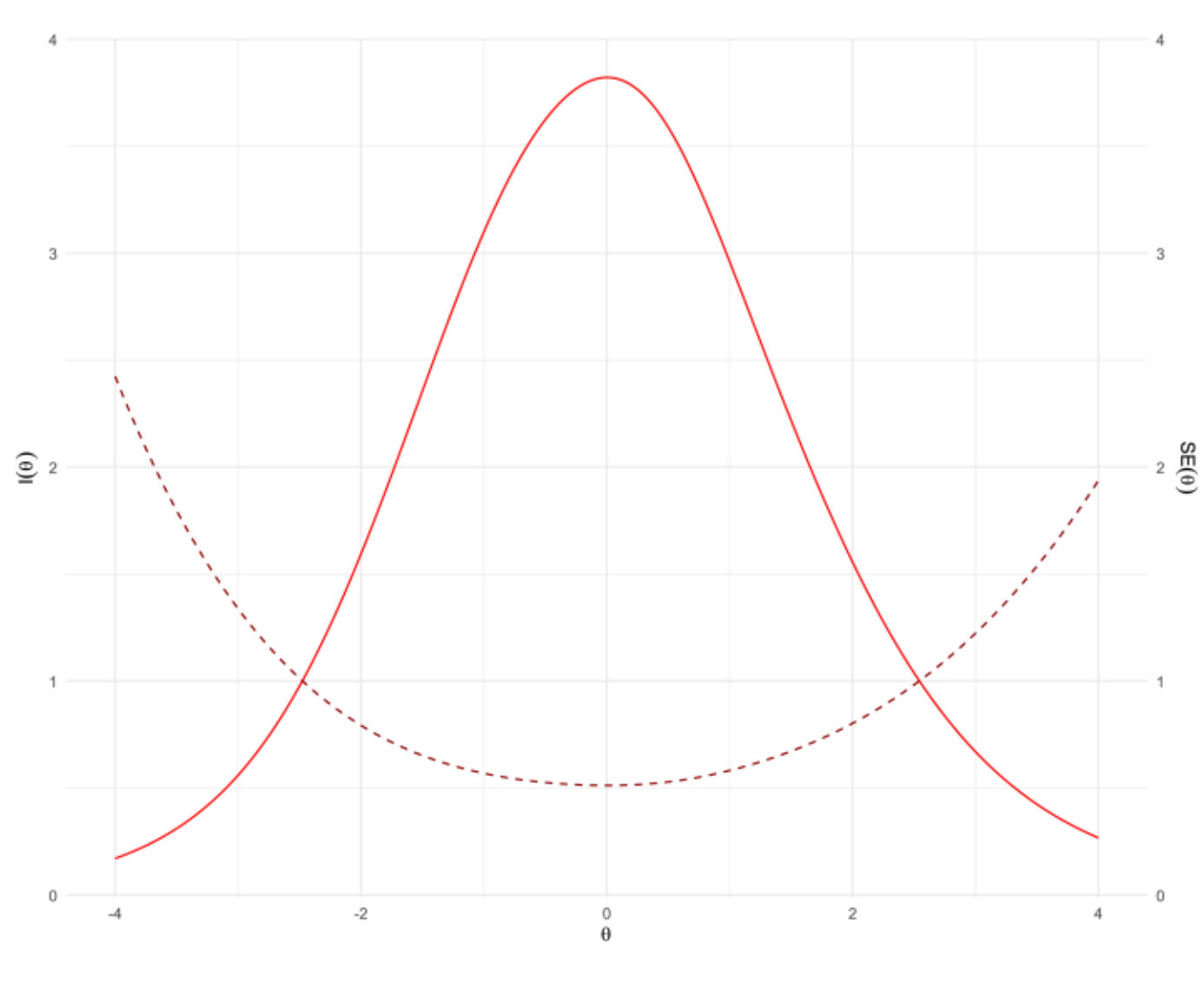

## 4 DISCUSSION

The present study developed the AI Competency Objective Scale (AICOS), an advanced instrument for measuring AI literacy that integrates current AI literacy and measurement models and takes into account the latest developments in the field of artificial intelligence by incorporating Generative AI Literacy (Annapureddy et al., 2024; Carolus, Koch, et al., 2023; Ng et al., 2021). Rather than developing an isolated test, the AICOS combines the strengths of content-validated scales to provide a more precise and nuanced measure of AI literacy. With 51 multiple-choice items, the test ensures efficient implementation and standardized evaluation. IRT analyses were used to determine item difficulty and to assess measurement accuracy across ability levels (Embretson & Reise, 2013). The results show that the AICOS measures with high precision, especially in the middle ability range ($\theta = 0.47$), while acceptable to good reliability is also achieved in the lower and higher ability levels (Table A7). The high variability of the item difficulties indicates that the test can measure a broad spectrum of ability levels in a differentiated manner (see Table 9 and Figure 3). Several quality criteria support the psychometric quality of the instrument. These include the analysis's confirmation of unidimensionality, the exclusion of local dependencies, and reliable item parameters and infit/outfit values. The instrument has been extensively validated, encompassing convergent, discriminant, and criterion validity, to address the limitations of existing measures. The CFA demonstrates high structural validity, confirming the test's suitability for measuring AI literacy. Convergent validity is demonstrated by a strong correlation between participants' ability scores and the H&P items measuring AI Literacy. Low, non-significant correlations with the MAILS support discriminant validity and demonstrate that subjective and objective AI Literacy are separate constructs (Carlson et al., 2009; Dunning, 2011; Klerck & Sweeney, 2007; Ma & Chen, 2023; Moore & Healy, 2008; Pinski et al., 2024; Waters et al., 2018). Several findings support criterion validity: Participants with a computer science background scored significantly higher on the test, demonstrating the test's ability to capture domain-specific knowledge differences. Additionally, high scores on the AICOS correlate with higher scores on the subsequent computer science quiz, confirming the test's predictive validity. With these comprehensive validity analyses and their implementation in German, the AICOS fills a key research gap. In contrast to many existing tests of AI literacy, which are based on self-assessment, the AICOS provides a more objective and empirically based assessment of AI literacy and thus expands the range of instruments for measuring AI literacy (e.g., Carolus, Koch, et al., 2023; Laupichler et al., 2023; Wang & Chuang, 2024). The scale measures the subscales Apply AI, Create AI, Detect AI, AI Ethics, Generative AI, and Understand AI. Depending on requirements, individual subscales can be selected and assessed in a focused manner, making the test flexible for use in broad and targeted studies. Additionally, an 18-item short version of the AICOS has been developed, yielding satisfactory results in terms of reliability and validity. Only the criterion validity of $r = 0.39$ is slightly below the threshold of $r = 0.40$. However, the selection of items for the AICOS-SV shows further potential (see Supplemental Material 4), which should be tested for validity and reliability in future studies on an independent sample. The short version enables efficient and economical screening for assessing AI competence.

The test was administered to a broadly representative sample with a gender balance of 53.74% male and 46.26% female participants. The mean age is 32.90 years. In contrast to other studies, such as Hornberger et al. (2023), the present sample primarily consists of individuals without student status who are employed full-time or part-time. The occupational background is diverse, with a focus on computing and administration. Most respondents have a high level of education, with a large proportion of people holding a German Abitur or Bachelor's degree. According to the German government's Demography Portal (2024), most Germans between the ages of 25 and 44 have an Abitur (comparable to a high school diploma), which is largely reflected in the study sample. The respondents rate their general understanding of AI and their use of AI in everyday life and at work as average. The study thus overcomes the limitations of Pinski et al. (2023), which focused exclusively on non-experts. Overall, the sample represents a heterogeneous German-speaking population, making the test suitable for comparison across different population groups. The test results revealed interesting demographic differences in gender, education, and professional background. Male participants scored higher in AI literacy. This finding is consistent with studies using subjective tests to measure AI literacy (see Asio, 2024; Toker Gokce et al., 2024), suggesting that gender differences are based not only on subjective perceptions but also on actual differences in ability.

Participants from technical and economic professions performed better on the test than those from social, creative, and service-oriented professions, which is consistent with the findings of Hornberger et al. (2023), who found similar results for technical academic backgrounds. Furthermore, the results show that individuals with academic degrees have higher AI literacy scores, a finding also supported by Lund et al. (2023), who observed a positive association between higher education levels and better data literacy, as well as greater interest in using AI. This can be attributed to better access to advanced educational opportunities and highlights the need to improve access to education for those with lower educational backgrounds.

### 4.1 Practical Implications

With its extensive validation and psychometric robustness, the AICOS is a versatile tool for research and practice, suitable for screening surveys, group comparisons, and skills assessments that inform targeted educational interventions, particularly in adult education. Its modular six-factor structure allows for the focused measurement of specific domains, such as AI ethics or generative AI, making it applicable to both broad and specialized contexts. For educators and policymakers, AICOS offers a reliable way to assess AI literacy across diverse populations, identify knowledge gaps, and inform curricula or training programs. In professional settings (e.g., technology and IT-related fields), it supports human resource development by evaluating employee competencies and guiding targeted upskilling. The instrument is available in two versions: a short form for efficient screening and a long form for more in-depth assessment. Although the tool was designed to apply to a broad adult population, including individuals with diverse educational and professional backgrounds, AI literacy needs may still vary among them. This diversity presents a challenge for developing a universally accessible instrument, and future adaptations may be necessary to ensure inclusivity across different user groups. Therefore, future adaptations of AICOS could incorporate tailored item content or alternative formats to better address these varied needs. Furthermore, integrating practical tasks or open-ended questions may enhance the assessment of applied AI skills beyond theoretical knowledge. As AI technologies, particularly generative AI, continue to evolve rapidly, AICOS remains a future-proof and adaptable instrument, capable of being updated to reflect emerging developments. Positioned within the broader context of AI literacy, AICOS makes a meaningful contribution to societal efforts to prepare individuals for an AI-driven world by providing a robust and scalable measure to support education policy and workforce development strategies.

### 4.2 Limitations and Future Research

Although the sample is heterogeneous and more representative than those in comparable studies, its high proportion of students and underrepresentation of older adults may limit the generalizability of the findings to the broader population. The broad age range (18–74) was unevenly distributed, and several occupational fields were sparsely represented. While our analyses did not reveal a significant age-related effect on AICOS performance, this may reflect the limited number of older participants rather than the absence of a true effect. Prior research on age differences in AI literacy is inconclusive, with some studies reporting no systematic effects (Laupichler et al., 2024; Mansoor et al., 2024), while others indicate that older adults have lower AI experience and usage (Kaur & Chen, 2023; Shandilya & Fan, 2022). Given these mixed findings, future studies should aim for more demographically balanced samples to explore potential differences in AI literacy across various life stages (e.g., students, professionals, older adults) and better capture the diversity of the general population. Furthermore, a selection bias cannot be ruled out, as participation was voluntary, which may have motivated individuals with a strong interest in AI. Additionally, the financial compensation provided an incentive that may have attracted certain groups to participate. A future research topic could be the standardization of the AICOS by establishing reference values for different target groups, such as age groups, educational levels, or professional groups, to improve the comparability of the results and the applicability of the test in various contexts. In this sense, as the AICOS is currently only available in German, future studies should aim to validate the instrument in additional languages, improving its international applicability and comparability.

In addition, the rapid development of AI could be a limitation, as some items in the test could become obsolete due to new technological advances or paradigm shifts. This would require periodic revisions of the scale. The AICOS measures only knowledge of AI, which means that AI literacy as a competency may be incompletely captured. In addition, knowledge is often weakly correlated with actual behavior (Ajzen et al., 2011), necessitating the development of behavioral measures to capture practical AI use more accurately. Future studies could complement knowledge-based measures with practical tasks, providing additional opportunities to validate the scale. Another potential problem is the different wording of the items. While some are factual and directly worded (e.g., GA10: "What is a 'prompt'?"), others are more application-oriented and include case examples (e.g., EA05, AA08). These formats may address different cognitive processes, as application-oriented tasks require additional skills, such as transferring knowledge to concrete scenarios. This could lead to biases depending on individual competencies. In addition, personalized elements within items (i.e., names such as "Samantha" in AA08) could elicit unintended cognitive or emotional reactions in participants that influence their responses, for example, if they associate them with real people they know. Further research should investigate the extent to which cognitive demands, influenced by item format and personalized elements, impact test scores. Future studies should incorporate pilot testing, complemented by qualitative methods (e.g., cognitive interviews, focus groups, or think-aloud protocols), to support these efforts. These approaches provide valuable insights into how participants understand and engage with the items. Gathering qualitative feedback helps identify problematic items and guide revisions, ultimately enhancing the instrument's quality before wider application.

Although our sample size is generally adequate for applying a 3PL IRT model (e.g., DeMars, 2010; Embretson & Reise, 2013), larger and more diverse samples would improve parameter stability, particularly for estimating the guessing parameter (e.g., De Ayala, 2013; Feuerstahler, 2022). Future research could incorporate pre-post reliability assessments to evaluate temporal stability, which would be particularly informative for items with lower discrimination. To further evaluate the scale's psychometric properties, item characteristics were examined in the 3PL model. Item and person parameters are theoretically separable, and the observed fit indices suggest that this assumption is reasonably met (De Ayala, 2013). Future studies could further investigate invariance using subgroup analyses or DIF methods. Within this framework, most items demonstrated adequate discrimination. However, a few items (e.g., DA08, GA11, UA11, UA18) exhibited relatively low discrimination ($a < 0.35$). These items were retained due to their content relevance, acceptable performance on other parameters, and prior validation (e.g., UA11), as well as because the overall reliability and validity of the scale remain high despite a few lower-discrimination items (De Ayala, 2013). However, the final version of AICOS was reduced from 107 to 51 items, which shortened the processing time and possibly reduced the cognitive load on participants. Despite the randomized presentation of the items and the largely controlled contextual effects, it cannot be ruled out that the excluded items may have influenced the perception and processing of the remaining items. Future studies should, therefore, examine the extent to which reducing the number of items might affect the validity of the results. Based on the application of the IRT methodology, the test's validity can be considered robust, as the removal of items did not bias the ability estimates for the remaining items (Embretson & Reise, 2000; Hambleton et al., 1991). Nevertheless, the reliability and validity of both the long and short versions should be further tested to confirm their measurement accuracy. For the long version, CFA indicates that AICOS can be used modularly, with the six-factor structure showing acceptable fit. Although the multidimensional model fits slightly better than the one-factor model, the improvement is minimal, supporting a unidimensional IRT approach for reasons of parsimony (Kline, 2023). Future studies could explore a multidimensional structure using a multidimensional IRT (MIRT), but this was not conducted here because the sample size was insufficient for stable parameter estimation and the marginal gain in fit does not justify the added complexity (Hambleton et al., 1991; Kline, 2023; Kose & Demirtasli, 2012; Tabachnick & Fidell, 2019). Regarding reliability, all subscales exceeded the acceptable threshold of .60 for instruments in the early stages of development (Hair et al., 2014), but not all reached the conventional benchmark of .70. From a psychometric perspective, this indicates that the items within certain subscales may not yet fully capture a coherent latent construct, underscoring the need for further refinement and item development. Aligned with the reliability results, some items exhibit low factor loadings, which might suggest exclusion under conventional criteria. However, they were retained due to their theoretical relevance, coverage of important construct facets, and positive loadings, thereby preserving content validity (Brown, 2015; DeVellis, 2017; Kline, 2023).

Importantly, fit indices indicate that the factor structure remains psychometrically sound, with most items meeting recommended thresholds (Brown, 2015; Comrey & Lee, 2013). Previous research also shows that lower factor loadings in large samples can still be estimated reliably and contribute to validity despite their small effect sizes (MacCallum et al., 1999). Nevertheless, these results suggest that the scale could be reviewed and, if necessary, revised in future studies to further optimize factor loadings. Concerning the short version, its current measurement accuracy represents an initial step that may guide further refinements and adaptations. Additionally, exploring different test configurations offers valuable directions for future research.

### 4.3 Conclusion

The development of the AICOS represents a significant step toward a valid and comprehensive measure of AI literacy. The test stands out for its methodological strengths, including comprehensive content coverage, a solid theoretical foundation, and rigorous validation based on scientific standards. Its inclusion of generative AI literacy, a dimension often overlooked in other instruments, further enhances its value. The AICOS also has the unique advantage of being administered to a heterogeneous sample, ensuring comparability across populations. This makes it a valuable tool for research and practical applications, particularly in educational and professional settings. Future studies should focus on validating and standardizing the AICOS in different contexts, including international ones, to further improve the comparability of results. Overall, the AICOS provides a solid foundation for advancing the measurement of AI literacy and supporting targeted education and training initiatives in an increasingly technology-driven society.

## 5 DECLARATION OF AI AND AI-ASSISTED TECHNOLOGIES IN THE WRITING PROCESS

During the preparation of this work, the authors utilized DeepL to enhance the readability and language of individual sentences. After using this tool, the authors reviewed and edited the content as needed and took full responsibility for the publication's content.

## 6 DECLARATION OF COMPETING INTEREST

The authors declare that they have no known competing financial interests or personal relationships that could have appeared to influence the work reported in this paper.

## 7 ETHICAL STATEMENT

The procedure performed in this study was conducted in accordance with the ethical standards outlined in the 1964 Declaration of Helsinki and its later amendments. An ethical review and approval were not required for the study of human participants in accordance with institutional requirements. The study did not involve medical, biological, personal, or sensitive data requiring specific ethical approval. All standard procedures to minimize risks for participants were strictly followed, including adherence to ethical guidelines for data collection. Informed consent was obtained from the participants to participate in the current study.

## 8 SUPPLEMENTARY MATERIALS, APPENDIX AND DATA AVAILABILITY

Supplementary material (including AICOS items) and appendices are available in the published version of this article: